\documentclass[notoc]{JHEP} % pas draft
\usepackage{amsmath}
\usepackage{epsfig}
\usepackage{amssymb,amsfonts}

\def\eps{\varepsilon}
\def\Z{\mathbb{Z}}
\def\RP{\mathbb{RP}}
\def\R{\mathbb{R}}
\def\O{{\cal O}}
\def\K{{\cal K}}

\title{Orientifolds of  the 3-sphere}
\author{Constantin  Bachas$^1$,  Nicolas Couchoud$^{1,2}$ 
 and Paul  Windey$^{2,3}$ \\
\begin{tabular}{ll}
$^1$ & Laboratoire de Physique Th{\'e}orique 
       de l'{\'E}cole Normale Sup{\'e}rieure
       \thanks{Unit{\'e} mixte du CNRS et de l'{\'E}cole Normale
       Sup{\'e}rieure, UMR 8549.} \\ 
     & 24  rue Lhomond, 75231 Paris cedex 05, France\\ \\
$^2$ & Laboratoire de Physique Th{\'e}orique et Hautes {\'E}nergies
       \thanks{Unit{\'e} mixte du CNRS et de l'Universit{\'e} de
       Paris VI et Paris VII, UMR 7589.}\\
     & Universit{\'e} Pierre et Marie Curie,  Paris VI\\
     & 4 place Jussieu, 75252 Paris cedex 05, France\\ \\
$^3$ & Theory Division, CERN\\
     & CH 1211, Geneva 23, Switzerland.
\end{tabular}
}

\abstract{
We study the geometry of orientifolds in  the  $SU(2)_k$  
WZW  model.
They correspond to the two
inequivalent,  orientation-reversing involutions of $S^3$, whose
fixed-point sets are:  the north and south poles (${\O}0$), or  the
equator two-sphere (${\O}2$).  We show how the geometric action 
of these  involutions  leads unambiguously 
to the  previously obtained  algebraic
results for the Klein bottle and
M{\"o}bius amplitudes. We give a semiclassical 
derivation of the selection rules and signs in
the crosscap couplings, paying particular
attention to discrete $B$-fluxes. 
 A novel observation, which does not follow
 from consistency of  the
 one-loop vacuum diagrams, is that 
 in the case of
the ${\O}0$ orientifolds only integer- or only
half-integer-spin Cardy states may  coexist.}

\preprint{LPTENS-01/39 
\\LPTHE-01-60
\\CERN-TH/2001-295  
\\hep-th/0111002}

\begin{document}
\section{Introduction and summary}

   The purpose of the present work is to study orientifolds of the
Wess--Zumino--Witten model on the group manifold of $SU(2)$. The problem
has been considered  from an algebraic point of view in
\cite{PSS,PSS2,SS,Sche}.  Here we will  elucidate  its geometry,
thereby clarifying and extending the previously obtained algebraic results.
Our  approach will be analogous to the one used for 
the  D-branes of this model in  \cite{BDS} (for some related works see
\cite{KS,AS,S,ARS,FFFS,Paw,MMS}).  
One important motivation for studying the $SU(2)$ WZW model
 is its  relevance in describing  the near-horizon geometry of
the Neveu--Schwarz fivebranes \cite{CHS}.\footnote{Early discussions
of orientifolds of this geometry can be found in \cite{Forste,BS}.}
The WZW  models  are, also,  of interest  as 
toy models  of warped compactifications of type-I
string theory (for a discussion of this point, and more references, 
 see \cite{Bstr}).

 Our results in this paper can be summarized as follows: 
\begin{enumerate}

\item The  possible
(classes of) orientifolds correspond to the
inequivalent orientation-reversing $\Z_2$ isometries of $S^3$. 
The requirement of orientation reversal follows from the invariance
of the Wess--Zumino term in the $\sigma$-model action. 
The fixed points of the two such inequivalent
 isometries  are (a) the north and south poles,
or (b) the equator two-sphere (denoted respectively
 ${\O}0$ and ${\O}2$).
\item The action of the $\Z_2$ isometries on closed-string vertex
operators dictates the form of the Klein-bottle
amplitudes in the direct channel. One recovers the
result  proposed in  \cite{PSS,PSS2,SS}. The same amplitudes in
 the transverse channel give the couplings of the crosscap to
closed-string states,  up to a sign.

\item In general, perturbative orientifolds can be further distinguished
by the $B$-flux they support  \cite{GimP,orB1}. This flux  is either
integer (${\O}^-$) or half-integer
(${\O}^+$). In our case, the two ${\O}0$'s  at the $S^3$ poles
are  of the same or opposite type,  according to the  
 parity of the Kac-Moody level $k$. 
This is consistent  with previous observations \cite{e1,e2,orB2,orB3}
 that an   ${\O}6$ orientifold 
changes type when intersecting a NS fivebrane.
This fact,  together with the
explicit form of hyperspherical harmonics,  explain furthermore   the selection
rules  found in the  crosscap couplings, 
and fix completely the sign ambiguities.

\item Combining the  crosscap  couplings
with the well-known D-brane couplings, gives
the M{\"o}bius-strip amplitudes in the closed channel. The
transformation
to the direct channel fixes the signs of the projections on
open-string states. We explain why these are consistent with the
geometric  action of the $\Z_2$ isometries. One subtle feature, that 
does not follow from consistency of  one-loop vacuum amplitudes alone,
is that only half of the WZW branes may  coexist for a given
choice of ${\O}0$ orientifolds.  

\item  We verify  that the  absolute value of the crosscap couplings
reduces,  for  large radius,  to the expected tensions of flat-space 
orientifolds. We also extend the analysis in \cite{BDS}, and  show that 
the Dirac-Born-Infeld action gives  the exact
(ratios of) D-brane couplings to all higher, closed-string harmonics
on $S^3$. Finally, we    discuss  briefly the extension 
of our analysis to $AdS_3$.    
\end{enumerate}

{\it Note added in proof}:
While we were completing  this paper, there appeared
the preprints  \cite{Br,HSS} which contain  some overlapping  results.

\boldmath
\section{Geometric ${\O}0$  and ${\O}2$ orientifolds}
\unboldmath
\label{secgeom}
The $SU(2)$ group manifold can be parametrized as
\begin{equation}
g  = \frac{1}{L}
\left(
\begin{array}{lll}
X_1+iX_2 &\quad & X_3+iX_4\cr
-X_3+iX_4 &\quad & X_1-i X_2\cr
\end{array}
\right) ,
\label{mat}
\end{equation}
with the $X_i$ taking values on a three-sphere of radius $L$. 
Two  standard parametrizations of the sphere are
(a) in terms of   polar coordinates: 
\begin{equation}
 X_1 = L\; {\rm cos}\psi\ ,\ \ 
X_2 = L\; {\rm sin}\psi\; {\rm cos}\theta\ , \ \  
X_3+iX_4 = L\; {\rm sin}\psi\; {\rm sin}\theta\; e^{i\phi}\ ,  
\end{equation}
with $\psi,\theta\in [0,\pi]$ and $\phi\in [0,2\pi]$, 
or (b) in terms of Euler angles:
\begin{equation}
X_1+iX_2 =  L\; {\rm sin}\alpha \;  e^{i\beta}\ , \ \ 
X_3 +iX_4 = L\; {\rm cos}\alpha \;  e^{i\gamma}\ ,
\end{equation}
with $\alpha\in [0,\pi/2]$ and  $\beta, \gamma\in [0,2\pi]$.
Strictly-speaking,
the Euler angles are  $\alpha$ and the two linear
 combinations $\beta\pm \gamma$.

  The orientifold operation, $\Omega h$,    is a combination of
worldsheet orientation reversal, which for closed strings reads
\begin{equation}
 \Omega :\   \sigma \to 2\pi -\sigma\ , \ \ {\rm or}\ \ 
 \ z \equiv e^{\tau +i\sigma} \to \bar z \ ,
\label{flip}
\end{equation}
and of   a  $\Z_2$ isometry $h$   
of  the target  manifold.  All   isometries  of the three-sphere can be
realized as $O(4)$ rotations in the embedding four-dimensional 
 space, so that 
modulo  conjugation by a group element
\begin{equation}
 h  =  {\rm diag} ( \pm 1, \pm 1, \pm 1, \pm 1 )\ .  
\label{a}
\end{equation}
The fact that $h$ is an isometry ensures the  invariance of  the metric term
in the $\sigma$-model action. However, the
non-trivial Neveu-Schwarz antisymmetric background 
imposes  one  extra   condition. Since $\Omega$ 
flips the orientation of any 
3-manifold  whose boundary is the string worldsheet, $h$
must  {\it flip the orientation
of the target three-sphere}  whose volume form is proportional
to the  NS field strength $H=dB$. 
This is required for the  invariance of
the Wess-Zumino term in the $\sigma$-model action. 
Put differently, we must make sure that the
orientifold projection does not eliminate
the $B$-field components that are turned on in the WZW background.
We are thus left   with two inequivalent  possibilities (see Figure 1):

\begin{itemize}
\item
${ h}_{0} =  {\rm diag} ( \; 1,  -1,  -1,  -1 )$ 
which maps    $g \to  g^\dagger$.
In terms of   polar and  of  Euler coordinates this action  reads:
\begin{equation}
(\psi, \theta, \phi) \to (\psi, \pi- \theta, \pi+\phi)\ , 
\ \ {\rm and}\ \ 
(\alpha, \beta, \gamma) \to (\alpha, -\beta, \pi+\gamma)\ . 
\label{o0}
\end{equation}
The fixed points of this  transformation  are 
 a pair of  ${\O}0$  orientifolds,   located 
at the north and at the south pole of the three-sphere (at $\psi=0,\pi$).\\
\item
$h_{2} =  {\rm diag} ( -1, \; 1, \; 1,  \; 1 )$ which maps 
$g \to -g^\dagger$ , or  explicitly
\begin{equation}
(\psi, \theta, \phi) \to (\pi-\psi, \theta, \phi)
\ \ {\rm and}\ \ 
(\alpha, \beta, \gamma) \to (\alpha, \pi-\beta, \gamma)\ .
\label{o2} 
\end{equation}
The fixed-point locus of this isometry  
is   an  ${\O}2$  orientifold that wraps  the equator two-sphere
(at $\psi=\pi/2$). 
\end{itemize}
Notice that there are neither   ${\O}1$, 
 nor   ${\O}3$ orientifolds.
Note also that in a complete  string background
  the  orientifolds may  acquire  
extra spatial  dimensions, as we will discuss later in the context
of the type-II NS fivebrane.

\vskip 1.2cm 

\begin{figure}[ht]
 \hskip 1.5 cm
       \hbox{\epsfxsize=130mm%
       \hfill~% Here be picture.
       \epsfbox{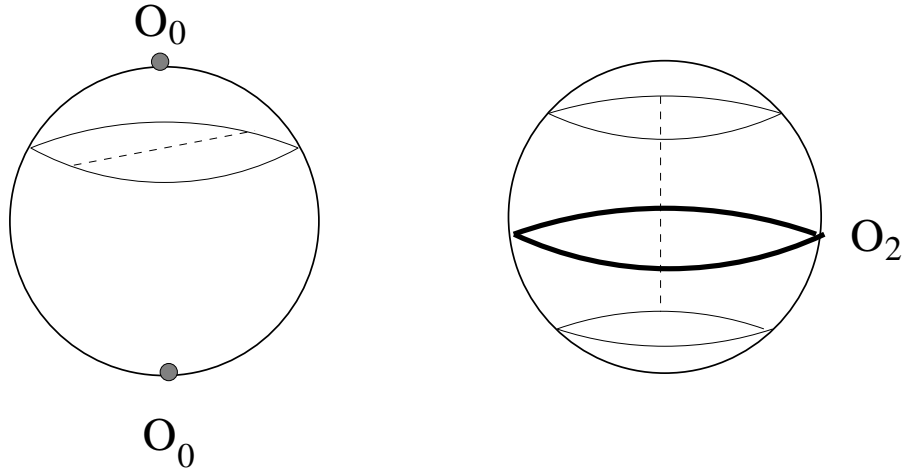}
       \hfill~}
       \caption{The two possible orientifolds for
 a three-sphere that is threaded by non-vanishing 
NS three-form flux. The
thin circles are two-spheres at fixed polar  angle
$\psi$,  while the broken
lines connect  pairs of  identified points. The ${\O}1$ and 
${\O}3$ orientifolds are not consistent with the background flux.}
\end{figure}

  In  general, the  type-II orientifold  may  contain both $\Omega h_0$
and $\Omega h_2$, and hence also their product $h_0h_2=-1$.
The latter is the freely-acting isometry that  is modded out when
passing  from $SU(2)$ to $SO(3)\simeq SU(2)/\Z_2$. 
In this case, quantization of the NS flux requires that the Kac-Moody level $k$
be  even.
 The corresponding orientifolds are,  
 in the language of \cite{PSS,PSS2,SS},
  `open descendants' of the non-diagonal  D-series models. 
The $SO(3)$ model in particular will be discussed in a separate  paper by one of us
\cite{cou}. 
More general orientifolds gauge a group of the form
 $ \Gamma \oplus \Omega h_0 \Gamma$ 
or $ \Gamma \oplus \Omega h_2 \Gamma$, 
 with $\Gamma$ any  discrete
group of orientation-preserving isometries. This would be relevant, 
 in particular, for orientifolds of the E-series models,
 as well as for orientifolds of string compactifications on 
the Lenz spaces $S^3/\Z_n$. We will not discuss such  models in the
present work.

%%%%%%%%%%%%%%%%%%%%%%%%%%%%%%%%%%%%%
%%%%%%%%%%%%%%%%%%%%%%%%%%%%%%%%%%%%

\section{Klein bottle and selection rules}
\label{klein}

 From the geometric actions \eqref{o0} and \eqref{o2} 
we can easily deduce
the Klein bottle amplitudes. These  implement  
 the orientifold projection 
on closed-string states, which in  the 
 (diagonal)   bosonic $A$-series models 
are  of the form
\begin{equation}
 {\cal P}(J^a_n, {\bar J}^{\bar a}_{\bar n})\;
  \vert j,m,\bar m\rangle \otimes
\; \vert {\rm rest}\rangle\ .  
\label{states}
\end{equation}
Here  ${\cal P}(J^a_n, {\bar J}^{\bar a}_{\bar n})$ is a polynomial in
the
left  and right  Kac-Moody currents, and  
$\vert j,m,\bar m  \rangle$ is a  lowest-weight state
 transforming  in the $(j,j)$ representation of the 
global  $SO(4)\simeq SU(2)_L\times SU(2)_R$. The 
allowed values of $j$  are 
 $0, 1/2, \cdots k/2$, and
$m,\bar m$ are the $J^3$ and $\bar J^3$ eigenvalues.
 The full string-theory
 background has  extra CFT components
besides the WZW model --  we have denoted the corresponding 
component of the closed-string state  by 
$\vert {\rm rest}\rangle$. Physical states must of course also obey the
Virasoro conditions.

We need the action of the orientifold operations
on \eqref{states}. First, notice
 that both $\Omega h_0$ and $\Omega h_2$  exchange 
 the left-  with  the right-moving  currents, 
\begin{equation}
\Omega h:\  J =\; \frac{k}{2}\;
 g \partial g^\dagger \;\; { \longleftrightarrow}
\;\; {\bar J} =\;\frac{k}{2}\;   g^\dagger {\bar \partial}  g\ .
\end{equation}
The two inequivalent orientifolds differ, however, in their action
 on  the primary states. The corresponding vertex operators are the
(ultra)spherical harmonics, which can be expressed as homogeneous
polynomials of degree $2j$ in the Cartesian coordinates $X_M$. 
In the Euler parametrization they take the form
\begin{equation}
{\cal D}^j_{m\bar m}(\alpha,\beta,\gamma)\; =
 \; e^{i(m+\bar m)\gamma}\;e^{i(m-\bar m)\beta}\;
P^j_{m\bar m}(\cos\alpha)\ ,  
\end{equation}
where   $P^j_{m\bar m}$ solve
a second-order differential equation, and are
related to the Jacobi functions \cite{vilenkin}.
 The   explicit form of these functions 
is not important  for our purposes here. All that matters is 
that they are symmetric under the interchange of $m$ and $\bar m$.
Using this fact, 
and the transformations
 \eqref{o0} and \eqref{o2},  we find the following  actions
on the  closed-string primaries:
\begin{equation}
\Omega h_0  \vert j,m,\bar m\rangle  =
  \;(-)^{m+\bar m} \vert j,\bar m,  m \rangle \; ,
\ \ \ {\rm and}
\ \ \ \   
\Omega h_2 \vert j,m,\bar m\rangle =  \; \vert j,\bar m,  m\rangle  \;. 
\end{equation}

We are  now  ready
to write down  the  Klein-bottle amplitudes. 
These  receive contributions only from $m=\bar m$ states. 
Denoting by $\chi_j$ the Kac-Moody characters, and using 
the  identity
$2m=2j\ (\rm{mod}\ 2)$,  we find:

\begin{equation}
\K_{(0)} = \; {1\over 2}\; {\rm Tr}\; \left[ \Omega h_0\; 
  q^{L_0+\bar L_0 -\frac{c}{12}}\right] \;
= \; {1\over 2}
\; \sum_{j=0}^{k/2}\;  (-)^{2j}\;  \chi_j( q^2)\; Z_{\rm rest}( q^2) \ , 
\label{klein0}
\end{equation}
and
\begin{equation}
\K_{(2)} = \; {1\over 2}\; {\rm Tr}\; \left[ \Omega h_2\; 
  q^{L_0+\bar L_0 -\frac{c}{ 12}}\right] \; 
 =\; {1\over 2}\;  \sum_{j=0}^{k/2}\;  \chi_j ( q^2)\; Z_{\rm rest}( q^2) \ .
\label{klein2}
\end{equation}
Here $ q = e^{- 2\pi  t}$, and $Z_{\rm rest}$ is the  contribution of
the CFT with which the WZW model is being  tensored. 
The  overall sign 
 must be 
 chosen so as  to symmetrize the states in the singlet ($j=0$) sector.
This ensures that  the identity operator is not eliminated by
the orientifold projection. 
Expressions \eqref{klein0} and \eqref{klein2}   agree
 with those proposed, on the basis of purely algebraic arguments,
 by Pradisi et al \cite{PSS}. 
We have derived them here from a  geometrical  viewpoint.

  To extract the tension and other properties of the orientifolds
we need to express the Klein bottle amplitudes  as an exchange 
of a closed string between  crosscaps. This transformation to the
`transverse channel' is achieved by 
the  change of variables
\begin{equation}
q= e^{- 2\pi  t} \rightarrow {\tilde q} = e^{-2\pi/ t}\ , 
\end{equation}
and by using  the  modular property  of the characters
\begin{equation}
\chi_i(q^2) = S_i^{\; j}\;  \chi_j (\sqrt{\tilde q}\;)\ . 
\end{equation}
Consistency requires the final result to be of the general form
\begin{equation}
 \K = \sum_{j=0}^{k/2}\;  (C^j)^2\;  \chi_j(\sqrt{\tilde q}\;)\;
 {\tilde Z}_{\rm rest}(\sqrt{\tilde q}\;)\ ,  
\end{equation}
with the 
$C^j$'s  giving  the orientifold (or crosscap) couplings  to  closed
strings in the $(j,j)$ representation of the current algebra.

The modular-transformation matrix of  the WZW model reads:
\begin{equation}
S_i^{\; j} = \sqrt{\frac{2}{k+2}}\;
 {\rm sin}\left(\frac{ (2i+1)(2j+1)\pi}{k+2}\right)\ . 
\end{equation}
Putting this expression 
in \eqref{klein0} and \eqref{klein2},  and doing the sums of the
sine functions,  leads to the   `crosscap coefficients' \cite{PSS}:
\begin{equation}
C^j_{(0)}\; =\; \eps^j_{(0)}\; 
   E_{2j+k}\; {\rm sin}\left(\frac{(2j+1)\pi}{2k+4}
\right)\;
\sqrt{{\cal N}_j}
\ ,
\label{cross0}  
\end{equation}
and
\begin{equation}
C^j_{(2)}\; =\;  \eps^j_{(2)}\; 
  E_{2j}\; {\rm cos}\left(\frac{(2j+1)\pi}{2k+4}
\right)\;\sqrt{{\cal N}_j} 
\ .
\label{cross2}  
\end{equation}
Here $E_n$ is a projector onto even integers (we will also need
the projector  onto odd integers, $O_n$). Explicitly:
\begin{equation}
E_n = \frac{1+(-)^n}{2}\ \ \ , \ \ \ O_n = \frac{1-(-)^n}{2}\ .
\end{equation}
We have also  introduced  the normalization factors
\begin{equation}
 ({\cal N}_j)^{-1}  \equiv  \sqrt{\frac{k+2}{2}}\;\;
  {\rm sin}\left(\frac{(2j+1)\pi}{k+2}
\right)\ , 
\label{norma}
\end{equation}
for reasons that  will become apparent later. 
Since we have only computed the squares of the $C^j$'s, there is
at this point a sign ambiguity parametrized by 
$\eps^j_{(0)}$ and $\eps^j_{(2)}$. 
 Notice that all  the
crosscap
couplings are real -- changing the overall sign of the Klein bottle
would have made them imaginary and is, hence, inconsistent.

   To understand the meaning of these formulae let us focus in
particular on the  couplings of the dilaton. These should
follow  from an  effective orientifold action
\begin{equation}
\label{oaction}
S = T_{\O} \int e^{\Phi} \sqrt{-\hat g} + \cdots\ ,
\end{equation} 
where $T_{\O}$ is the  orientifold tension, 
$\hat g$ the determinant of the induced  metric, and the dots stand
for higher $\alpha^\prime$ corrections. The power of $e^{\Phi}$
is the Euler character of the real projective plane, which is the same
as that of the disk diagram. Now let us expand the dilaton in 
 terms of hyperspherical harmonics:
\begin{equation}
\label{dilaton}
\Phi = \sum_j \; \sum_{L=0}^{2j}\; 
\sum_{ M=-L}^L \; \Phi_{jLM}\;  
F_{jLM}(\psi)\;
 Y_{LM}(\theta,\phi)\ .
\end{equation}
We are here using
 a  basis in which, instead  
of fixing the eigenvalues of $J^3$ and $\bar J^3$
 (as in the previous section), we are 
fixing the  angular momentum under the diagonal $SO(3)$ subgroup
of $SO(4)\simeq SU(2)\times SU(2)$. 
This is  convenient  because
only the $L=M=0$ components  couple to spherical
sources, like our  orientifolds or  the WZW D-branes.
The other components in the decomposition \eqref{dilaton} will vanish
upon integration over $\theta$ and $\phi$.
The coupling of the $j$th dilaton harmonic to a spherical source
located at  $\psi_0$ must, therefore, be  proportional
to $F_{j00}(\psi_0)$.

Consider first  the ${\O}2$ orientifold,  
which is located at the equator of the three-sphere, $\psi=\pi/2$. 
The functions $F_{j00}(\psi)$  are the Gegenbauer polynomials:
\begin{equation}
F_{j00}(\psi) = \sqrt{\frac{2}{\pi}}\; 
                \frac{{\rm sin}(2j+1)\psi}{{\rm sin}\psi}\ ,
\label{geg}
\end{equation}
while  $Y_{00}= 1/\sqrt{4\pi}$. 
From equations \eqref{oaction} and \eqref{dilaton} we thus derive the
following effective couplings of 
 ${\O}2$ to the  $\Phi_{j00}$~:
\begin{equation}
\label{effcouplings}
C_{{\rm eff}\ (2)}^{j} \; = \;  2 \sqrt{2}\;
 T_{{\O}2}\;  L^2 \;(-)^j\;  E_{2j}\;\ ,
\end{equation}
where $L$ is the radius of the three-sphere, and we have used the 
obvious identity~: 
${\rm sin}[(2j+1)\pi/2] = (-)^j\;  E_{2j}$ for integer or
half-integer $j$.

This effective field-theory calculation `explains', firstly,  the
selection rule, i.e. the presence of the projector
$E_{2j}$. Indeed,  
harmonics corresponding to odd $2j$ 
vanish on the equator
two-sphere, and hence do not couple to the 
 ${\O}2$ orientifold (at least to leading order in $\alpha^\prime$).
The same   calculation   
 fixes  also the sign ambiguity in 
expression \eqref{cross2}:
\begin{equation}
\eps^j_{(2)}=  \; (-)^j \; {\rm sign}(T_{{\O}2}) \ .
\end{equation}
We will confirm these signs  by  computing  the M{\"o}bius
amplitudes later on.
A more detailed comparison of 
 \eqref{effcouplings} with \eqref{cross2} 
is a priori possible only in the semiclassical (large radius)  limit, and 
requires a careful 
match of  the normalizations 
of  exchanged closed-string states. 
 We will return  to  this problem
 in the following section. Suffice, for the time being,
 to say that both  the selection rules, 
and  the sign of the
couplings is not affected by normalization factors.

 Let us consider next the ${\O}0$ orientifolds, whose coupling
to the dilaton is the sum of a north-pole and a south-pole term:
\begin{equation}
\label{eff0}
T_{{\O}0}^{\rm north} \Phi(\psi=0) + T_{{\O}0}^{\rm south} \Phi(\psi=\pi)\ . 
\end{equation}
Now  the only 
$F_{jLM}$ that don't  vanish at the poles are those with $L=M=0$ (or else
the function $\Phi$ would have been singular).
Evaluating   the Gegenbauer polynomials
at the poles (by taking a limit) leads to 
 the following effective ${\O}0$ couplings:
\begin{equation}
\label{effcouplings0}
C_{{\rm eff}\ (0)}^{j} \; = \;  \frac{ \sqrt{2}\;(2j+1)}{2\pi}\;
[\; T_{{\O}0}^{\rm north}\;  +\; (-)^{2j}\; T_{{\O}0}^{\rm south}\; ]\ . 
\end{equation}
To proceed further we next  need to look into the precise nature
of  the north and south  orientifolds. 
Indeed, as explained very clearly in   \cite{orB1} (see also
\cite{GimP,orB2,orB3}),  perturbative orientifolds come in two types
according to the  discrete torsion of  the Neveu Schwarz field B. 
\footnote{Torsion associated with the Ramond-Ramond fields is invisible
in perturbation theory, and will not concern us here.} 
More explicitly,  the orientifolding operation replaces
a small sphere $S^n$ around
 the orientifold by one  copy of  $\RP^n$.
In the case at hand, assuming for simplicity that the
remaining  space time coordinates are spectators, 
we have  $n=2$. The flux of $B$ through this $\RP^2$ can be  either 
half-integer or integer -- this is the only  gauge-invariant statement
 one can make.
The corresponding orientifolds are denoted   ${\O}^+$ and 
${\O}^-$. These  give rise,  in the flat-space limit,   
to  respectively symplectic or  orthogonal D-brane gauge groups.

   What kind of orientifolds do we have in our case? A priori we may think
we are free to choose,
 but there is one global consistency condition.
The B-flux out of the $\RP^2$ around a pole is  half
the flux coming 
 out of the corresponding
$S^2$ before the orientifold projection. Consider now two infinitesimal
two-spheres around the north and the south poles of $S^3$. Since
(in units $4\pi^2\alpha^\prime =1$) 
\begin{equation}
\label{ortype}
 \int_{\rm north} B  - 
\int_{\rm south} B   \; =\; 
 \int_{S3} H \; =\; k  \ , 
\end{equation}
the two orientifolds are of the { same type} for
$k$ { even}, and of { opposite type}  if $k$ is { odd.}
\footnote{Note that the usual shift $k\to k+2$,  that takes care of $\alpha^\prime$
corrections,  does not modify this conclusion since $2$ is even.}
Orientifolds of opposite type have  tensions that differ by a sign, 
so  that
\begin{equation}
 T_{{\O}0}^{\rm north} = (-)^k\;  T_{{\O}0}^{\rm south}\ .
\end{equation}
Plugging into
\eqref{effcouplings0}
 leads to our final expression for the effective ${\O}0$ couplings:
\begin{equation}
\label{effcouplings00}
C_{{\rm eff}\ (0)}^{j} \; = \;  \frac{ \sqrt{2}\;(2j+1)}{\pi}\;
 T_{{\O}0}^{\rm north}\; E_{2j+k}\ .  
\end{equation}
This  `explains' the selection rule 
found in \eqref{cross0} 
by the Klein-bottle calculation,  and fixes also the sign ambiguity:
\begin{equation}
\label{signn}
\eps^j_{(0)} = {\rm sign}\;( T_{{\O}0}^{\rm north})\ .
\end{equation}
The M{\"o}bius strip calculation
in section 5 will give an independent confirmation of these signs. 
First, however, we  turn to a  comparison of  the  orientifold couplings $C^j$ 
with  the couplings of  WZW D-branes.

%%%%%%%%%%%%%%%%%%%%%%%%%%%%%%%%%%%%%%%%%%%%
%%%%%%%%%%%%%%%%%%%%%%%%%%%%%%%%%%%%%%%%%%%
%%%%%%%%%%%%%%%%%%%%%%%%%%%%%%%%%%%%%%%%%%%

\section{Comparing D-branes and  orientifolds}

The WZW $SU(2)$ model has 
 one type of Cardy state \cite{Ca}
 for each integer or half-integer value of the spin $s$, with
$0\le s\le k/2$. The 
explicit expression of these boundary   states  gives directly their
couplings  to closed-string  fields.    
Equivalently, one can read these couplings off  the annulus  diagram.
For an open string
stretching between  a 
 D-brane of type $r$ and  one of type $s$ the annulus diagram reads:
\begin{equation}
{\cal A}_{rs} =  {\rm Tr}\;
 \left(  \sqrt{q}^{\; L_0 -\frac{c}{24}}
\right)  
=  \; \sum_{j=0}^{k/2} N_{rs}^{\ \ j}\;  \chi_j(\sqrt{q}\;)
 \; Z_{rest}(\sqrt{q}\;) \ .
\label{annulus}
\end{equation}
Here  $N_{rs}^{\ \ j}$ are the Verlinde fusion coefficients, whose
non-zero values are 
\begin{equation}
N_{rs}^{\ \ j} = 1 \ , \ \ {\rm for } \ 
\vert r-s \vert \ \le j \le \ 
{\rm min}(r+s,\; k-r-s)\ \ 
.  
\end{equation}
Using the modular transformation of the characters  we
can write  the  amplitude in  the transverse (cylinder) channel as follows: 
\begin{equation}
{\cal A}_{rs} = \sum_{j=0}^{k/2}\;  D^j_r D^j_s \; \chi_j({\tilde q}^2)\;
{\tilde Z}_{rest}({\tilde q}^2)\  , 
\end{equation}
where  
\begin{equation}
D^j_s =  \;  {\rm sin}\left(\frac{(2j+1)(2s+1)\pi}{k+2}
\right)\;
 \sqrt{{\cal N}_j}\ \ \  , 
\label{dbra}
\end{equation}
and ${{\cal N}_j}$ are the  normalization coefficients
 given
by eq. \eqref{norma}. The $D^j_r$ describe  the couplings
 of a  D-brane of type $r$ to the closed strings 
in the $(j,j)$ representation of the current algebra. 
They are the counterparts of the crosscap coefficients
$C^j$ of the previous section. Note that in an orientifold
background  they should be multiplied by an extra factor
 of ${1/\sqrt{2}}$.

The semiclassical study  of the Cardy states \cite{BDS} is based
on the  Dirac-Born-Infeld action for
  a D2-brane,\footnote{There is  also a dual description in terms of
a D-particle matrix model, see 
 \cite{ARS,DO4,DO3,DO2,DO1}.}      
\begin{equation}
\label{DBI}
S  = T_{D2}\;  \int \;
  e^{\Phi}\;  \sqrt{-{\rm det}\; (\hat g + \hat B + 2\pi\alpha^\prime F)}
\ + \cdots  \ . 
\end{equation}
Here $T_{D2}$ is the brane tension and $F$ the worldvolume gauge
field. 
We consider for definiteness seven  flat spectator dimensions,  so
that $g$ is the natural metric on  $R^7\times S^3$. 
This background has ofcourse a  dilaton tadpole, but this  will
not affect  our present discussion. 
Let us now choose   a convenient 
gauge  in which the Neveu-Schwarz background  reads 
\begin{equation}
\label{back}
B = \left[ L^2  \left( \psi 
 - \frac{\sin 2\psi}{2}\right)+ {\pi}{\alpha'} n_0\right]\;
\sin\theta\,  d\theta \, d\phi\ ,
\end{equation}
where $n_0$ is an integer parametrizing a  (residual)  freedom of  
`large' $\psi$-independent 
gauge transformations. We will say more about this freedom later on.
Using the underlying group symmetry, we can furthermore 
bring the  center-of-gravity of the branes  to the north pole.
The stable  configurations  can then be shown \cite{BDS} to correspond to
two-spheres   spanned by 
 the polar angles $\theta$ and $ \phi$, and  
 carrying  a  worldvolume `monopole' flux
\begin{equation}
 F = -{n\over 2}\; \sin\theta\,  d\theta \, d\phi\ .  
\end{equation}
The branes sit at the
latitude
\begin{equation}
\psi_n \;= \;(n - n_0)\;\frac{\pi\alpha'}{L^2}\ ,
\label{min}
\end{equation}
where the DBI energy is minimized. 
To make contact with the CFT, one must relate  the label
$s$ of the Cardy states to  the (gauge-invariant)
 flux, 
\begin{equation}
n-n_0 = 2s+1\ ,
\end{equation}
and identify the three-sphere radius with the  Kac-Moody level, 
$L^2 = (k+2)\alpha'$. We have included here
the well-known radius  shift, which 
effectively resums an entire  series of $\alpha^\prime$ corrections
to the WZW background.

Consider now the dilaton couplings, which follow by linearizing 
the DBI action \eqref{DBI}. Plugging  (4.6-4.8) in this action one finds
\begin{equation}
 \sqrt{-{\rm det}(\hat g + \hat B + 2\pi\alpha^\prime F)}
 = L^2\,  \sin\psi_n \, \sin\theta\ .
\label{db}
\end{equation}
Notice the single power of ${\rm sin}\psi_n$ in this expression.
Simple geometrical  area is  proportional to  ${\rm sin}^2\psi_n$,
but the DBI energy is larger because of the contribution of
$B$ and $F$. Expanding the dilaton in hyperspherical
harmonics as in \eqref{dilaton}, 
and using the 
Gegenbauer polynomials \eqref{geg},  leads easily to
the following  effective D-brane couplings:
\begin{equation}
 D^j_{{\rm eff}\ s}\; =\; 2\sqrt{2}
\; T_{D2}\; L^2 \, \sin\left[ (2j+1)(2s+1)\pi
\alpha'/L^2 \right]\ .  
\end{equation}
These should be compared to the exact CFT results  \eqref{dbra}.

  The first thing to notice is that  for given $j$, and after taking
into account the radius shift, the  ratios
of effective couplings match exactly those of the CFT,  
\begin{equation}
{D^j_{{\rm eff}\ s}\over  D^j_{{\rm eff}\ r}} = {D^j_s\over D^j_r}\ .
\end{equation}
This  agreement was pointed out for $j=0$ in \cite{BDS}, and we
see here that it continues to hold for all $j$. 
It suggests that the DBI and CFT calculations differ only in the
precise normalization of the exchanged closed-string states.
In order to  compare this normalization, we must use the relation
between the two different bases of hyperspherical harmonics, 
\begin{equation}
\vert j, L=M=0 \rangle = {1\over \sqrt{2j+1}} \; \sum_{m\; =-j}^j \vert j, m, -m \rangle
\end{equation}
The field theory calculation gives the coupling to the state on the
left-hand-side, while the CFT calculation gives the coupling to each
individual state $\vert j, m, -m \rangle$ . 
The meaningful ratio is thus 
\begin{equation}
\label{normalll} 
{1\over 2j+1}\; \left( {D^j_{{\rm eff}\ s}\over D^j_s}\right)^2 
\;  = \; 4\pi \sqrt{2}\; (T_{D2}\,\alpha^\prime)^2 \;
(k+2)^{3/2} \;  \left[ {\sin x_j  \over x_j }\right] 
\end{equation}
where
\begin{equation}
x_j \equiv {(2j+1)\pi\over k+2}\ .
\end{equation}
The $j$-independent factor in the right-hand-side of 
\eqref{normalll} is uninteresting --
it simply accounts for the gravitational coupling and $S^3$ volume sitting in front of
the 7d supergravity lagrangian \cite{BDS}. The remaining factor inside the  brackets 
approaches one in the semiclassical, $k\to\infty$,  limit. 
For finite $k$ it should, in principle, arise from $\alpha^\prime$ corrections to the closed-string
action. These combine  elegantly, in the case at hand,  to form  the ratio of a quantum and a
classical dimension. 
This fact could prove  of interest in searching for gravitational analogs of the Born-Infeld
action  (see  \cite{DG} for a  recent discussion
of non-linear gravitational actions).

  Let us  go back now  to the orientifolds, whose  closed-string couplings are
given by  equations
\eqref{cross0} and \eqref{cross2}. 
In section 3 we explained  the origin of the selection rules,
and reduced the sign ambiguities to an overall, $j$-independent sign.
In order  to better 
understand  now  the  magnitude of the crosscap couplings, 
it is convenient to 
 compare them  to those of  D-branes, whose normalization
we have  discussed with great care.  Taking ratios one finds~:
\begin{equation}
{C^j_{(0)}\; \over D^j_s} =  \pm E_{2j+k}\; {{\rm sin}{(x_j/ 2)}\over
{\rm sin}{[(2s+1) x_j}]}\ ,
\end{equation}
and
\begin{equation}
\label{lims}
{C^j_{(2)}\over  D^j_s} =  \pm E_{2j}\; {{\rm cos}(x_j/ 2)\over 
{\rm sin}{[(2s+1) x_j}]}\ . 
\end{equation}
In  the large-radius limit  
the D2-brane reduces to a collection of $2s+1$ D-particles,  all
sitting at the north pole of the three sphere. Using  
the well-known relation between D-brane tensions (see \cite{rev})
 one finds~: 
\begin{equation}
\label{footeq}
{C^0_{(0)}\; \over D^0_s}\to \pm {1\over 2(2s+1)} E_{k}\; \ \
{\rm and} \ \ {C^0_{(2)}\; \over D^0_s} \to  \pm {4\pi L^2 T_{D2}\over
  (2s+1)T_{D0}}\ . 
\end{equation}
This is what one should expect for, respectively,  
two ${\cal O}0$ orientifolds
 (of same or opposite type according to the parity of $k$),  and for an
${\cal O}2$ orientifold wrapping 
 the equator two sphere.\footnote{To check the factors
of $2$ in the above expressions,  
note for example that 
in the case of a flat three-torus and a
single D-particle ($s=0$), a similar  calculation
would have given a numerical factor $2^{3/2}/\sqrt{2}$. 
Since on $T^3$  there are eight  ${\cal O}0$s , while 
on $S^3$ there are only two, 
we have to  divide this  result by 4. This
 reproduces  the factor in \eqref{footeq}.}

  It is intriguing to observe
that  the exact expressions of the crosscap couplings of the
${\cal O}0$s, resemble  those for  a (hypothetical)
D-brane with (gauge-invariant) flux one half. 
This is suggestive of an exact
semiclassical argument, like the one that worked for the D-branes, 
though there  is no a priori reason why
 such an argument should exist.

%%%%%%%%%%%%%%%%%%%%%%%%%%%%%%%%%%%%%%%%%
%%%%%%%%%%%%%%%%%%%%%%%%%%%%%%%%%%%%%%%%%%%%%%%

\section{M{\"o}bius amplitudes}
\label{Mobius}

After the Klein bottle and the annulus, we turn now our attention
to 
the M{\"o}bius amplitude  which implements the orientifold projection on
open-string states. Our  geometric analysis will both  clarify  and
extend  the algebraic results of \cite{PSS}. 
We will show, in particular,  that
in  the ${\cal O}0$ case only half of the WZW D-branes can  coexist. 
This is a finer requirement, not dictated  
solely by  consistency  of the genus-one vacuum
amplitudes. 

 In order to write down M{\"o}bius amplitudes it is convenient
to work with the basis of real characters:
\begin{equation}
\hat{\chi}_j(q) \equiv e^{-i\pi(h_j-c/24)}\;\chi_j(-\sqrt{q})\ . 
\end{equation}
The M{\"o}bius amplitude is given by a
linear combination of $\hat\chi_j({q})$ in the direct channel, and
$\hat\chi_j({\tilde{q}})$ in the transverse channel.
The relevant  transformation is implemented by the  matrix 
\begin{equation}
P = T^{1/2} S T^2 S T^{1/2}\ .   
\end{equation}
In the case of  the WZW $SU(2)$  model a straightforward
calculation gives  \cite{PSS,PSS2}
\begin{equation}
\label{P}
P_i^{\ j} = \frac{2}{\sqrt{k+2}}\; 
        \sin\left(\frac{\pi(2i+1)(2j+1)}{2(k+2)}\right) E_{2i+2j+k}.
\end{equation}
We have included, for the reader's convenience,  the calculation
of this $P$ matrix (which requires some finite Gauss sums)
 in an appendix.  

 Since we only study $SO(3)$-invariant configurations, our
D2-branes will be always parallel  to the ${\O}2$, or have
the ${\O}0$s at their geometrical center. More general configurations,
including branes and orientifolds at angles,
are beyond our present scope. We  also restrict our attention to
elementary  branes, and will not discuss Chan-Paton multiplicities.

\boldmath
\subsection{D-branes of  ${\O}0$ models}
\unboldmath
Since $h_0$  maps the spherical D2-branes onto
themselves,  the M{\"o}bius amplitude involves  in this case 
open strings
attached to  a single D-brane. 
The corresponding
 primaries
are spherical harmonics with standard parity properties, 
\begin{equation}
Y_{lm}(\pi-\theta,\phi+\pi) = (-)^l\; Y_{lm}(\theta,\phi)\ .   
\end{equation}
For a D-brane of type $r$,  
the M{\"o}bius amplitude in the open channel therefore reads  
\begin{equation}
{\cal M}_{(0)\ r} =  \eps'_r 
\sum_{l=0}^{l_{0}} \;(-)^l\; \hat{\chi}_l(q)\; \hat
                Z_{\rm rest}(q)\ ,  
\end{equation}
where $l_{0}= {\rm min}(2r, k-2r)$, and we have allowed for 
an arbitrary overall  sign,    $\eps'_r$,
which would  determine   the type
of projection for   D-brane of type $r$. 
Transforming the above  amplitude
 to  the closed channel with the $P$ matrix, leads  after some
elementary algebra to:
\begin{equation}
{\cal M}_{(0)\ r} =  \eps'_r (-)^{2r}\; 
  \sum_j \;
\vert C^j_{(0)}\vert\;  D^j_r\; \hat{\chi}_j(\tilde{q})\; 
\hat Z_{\rm rest}(\tilde{q})\  . 
\end{equation}
The crosscap and D-brane coefficients are given by 
\eqref{cross0} and \eqref{dbra}. 
Consistency requires the $C^j_{(0)}$ to enter in this expression
without absolute value,    or extra signs. 
This is  the case provided: (a)  the 
sign  of the crosscap coefficients  does  not depend on 
the spin $j$, in agreement with our 
semiclassical reasonning leading to \eqref{signn} in  
section 3, and (b) the projection 
alternates with the D-brane type, 
\begin{equation}
\label{alternn}
 \eps'_r = {\rm sign}\;( T_{{\O}0}^{\rm north})\; (-)^{2r}\ . 
\end{equation}
Note that at the level of genus-one vacuum diagrams  all types
of D-branes are a priori allowed, provided the above
condition is respected.

   From the geometric viewpoint this conclusion cannot, however,
 be correct. The $l=0$ open-string states include three translation
zero modes \cite{BDS} that would survive the projection if
 $\eps'_r$ were positive. But  rigid  translations  displace
the geometric center of the brane away from the north pole
of $S^3$, and are inconsistent
with the $h_0$  symmetry. Thus $\eps'_r$ better be always negative.
Equation \eqref{alternn} then implies 
 that only branes of `integer type' are in the spectrum
for an ${\O}0^-$ orientifold at the north pole, while only
`half-integer types' are allowed if the north-pole orientifold is
${\O}0^+$. In either case,  there are no zero modes in the open-string
 spectrum.

  Another  argument,  leading to the same
conclusion,   goes as follows:  Since the
D2-branes in the orientifold theory are geometrically
$\RP^2$s, the monopole flux ($n$)  through them  
has to be even. From  eq. \eqref{min},  which relates
 $n$ to  the D-brane type, we conclude  that only D-branes of
integer (half-integer) type $s$ are   allowed when  $n_0$ is
odd (even). But $n_0/2$ is precisely the $B$-flux through an $\RP^2$
around the orientifold north-pole, as can be seen from eq. \eqref{back}.
Thus the type of the north-pole orientifold determines the
set of allowed Cardy states, in agreement with our previous 
argument about zero modes.

\boldmath
\subsection{D-branes in ${\O}2$ models}
\unboldmath
Let us move on now to 
the involution  $h_2$, which   maps a D-brane of type $r$ onto its 
mirror brane of type $k/2-r$. 
The D2-branes will thus come in pairs, except when $k$ is even
in which case there is an invariant brane at the equator.
The generic  M{\"o}bius amplitude involves strings  stretching
between  mirror branes,
and transforming in  Kac-Moody representations with
spin $j= k/2-2r, \cdots k/2$. 
Since the spherical harmonics do not transform under $h_2$,
we expect the $\Omega h_2$ projection to be 
$j$-independent.\footnote{This is obvious for the strings on
 equatorial brane, from which one
can then generalize to other branes by continuity.} 
 Thus the  
 M{\"o}bius amplitude in the open channel should read:
\begin{equation}
{\cal M}_{(2)\ r} = \; \eps'_r
                \sum_{j=k/2-2r}^{k/2} \; \hat{\chi}_j(q)
\; \hat Z_{\rm rest}({q})\  . 
\end{equation}
Transforming to the transverse channel with the $P$ matrix, and
comparing with the expected form, 
\begin{equation}
{\cal M}_{(2)\ r} = \sum_j 
                      C^j_{(2)} D^j_r\; \hat{\chi}_j(\tilde{q})
\; \hat Z_{\rm rest}(\tilde{q})
\ , 
\end{equation}
we find  that $\eps'_r = {\rm sign}\;( T_{{\O}2}) $ for all  $r$.
All (pairs) of WZW D-branes are allowed here, and  all
but the equatorial brane have zero modes. These  correspond to
simultaneous rigid motions  of a brane and of its mirror image.

Let us conclude this section with a subtle point. 
Naively, one may have thought that
the $B$ field should vanish on the equator two-sphere, where  the
geometric $h_2$ action is trivial. 
This is the case  for the  background
\begin{equation}
\label{Bnaif}
B = L^2  \left( \psi - \frac{\sin 2\psi}{2} - \frac{\pi}{2} \right)
   \sin\theta \,d\theta \,d\phi\ ,
\end{equation}
which 
corresponds  to the choice $-n_0 = k/2+1$
in eq. \eqref{back}. For $k$ even this is 
 indeed an allowed
choice, but when $k$ is odd it has (observable) 
singularities at the poles.  Does this mean that 
the ${\O}2$ orientifold is inconsistent unless $k$ is even
(like the $SU(2)/\Z_2$ orbifold)?
 The answer is  `no'. The field  $B$ can, in fact,  be non-zero on 
the  orientifold two-sphere, provided $B$ and $-B$ are
related by  a (large) gauge transformation.
 When $k$ is odd our spherical orientifold 
has  a non-trivial  $\Z_2$  flux on  its  worldvolume, which
is however perfectly consistent. 
An analogous situation is  known to arise
in the flat,  toroidal case \cite{BPS}. Notice that the discrete
$B$-flux on  the ${\O2}$ should not be confused with the
discrete fluxes out of an  $\RP^2$ surrounding an ${\O0}$.
The difference  is similar  to the difference between
a magnetic charge  and a flux line.

%%%%%%%%%%%%%%%%%%%%%%%%%%%%%%%%%%%%%%%%%%%%%%%%%
%%%%%%%%%%%%%%%%%%%%%%%%%%%%%%%%%%%%%%%%%%%%%%%%%%

\boldmath
\section{ $SL(2,\R)$ and NS fivebranes}
\unboldmath

  It would be interesting to extend our analysis to
GKO coset models, Gepner  models, and to other group manifolds.
In place of concluding remarks, let us briefly discuss here the
special case of the group manifold $SL(2,\R)$, and comment also
on the embedding of our orientifolds in the near-horizon
geometry of NS fivebranes.

The manifold of  $SL(2,\R)$ is a hyperboloid in $\R^{2,2}$. 
This is obvious  when one parametrizes  real $2\times 2$ matrices  as follows: 
\begin{equation}
g = {1\over L} \; \left( \begin{array}{ccc} 
    X^0+X^1 & \quad & X^2+X^3 \\ X^2-X^3 & \quad & X^0-X^1 
    \end{array} \right) \ .
\end{equation}
Cylindrical coordinates are defined through the relations
\begin{equation}
X^0 + iX^3 = L \cosh \rho\; e^{i\tau}\ ,
\quad X^1 + iX^2 = L \sinh \rho\; e^{i\phi}\ .
\end{equation}
The  $AdS_3$  spacetime  is the universal cover,  
obtained  by ignoring the periodic identification
of  the time  coordinate $\tau$.

We are interested in orientation-reversing $\Z_2$ isometries.
Because of the indefinite metric there exist 
four inequivalent such elements of
$O(2,2)$, as opposed to only two for $O(4)$. 
They can be chosen as follows: 
\begin{itemize}
\item $X^2\to -X^2$ , or $(\tau, \phi) \to (\tau, - \phi)$ ;
\item $X^3\to -X^3$ , or $(\tau, \phi) \to (- \tau, \phi)$ ;
\item $X^{1,2,3}\to -X^{1,2,3}$ , or $(\tau, \phi) \to (-\tau, \pi+ \phi)$ ;
\item $X^{0,2,3}\to -X^{0,2,3}$ , or $(\tau, \phi) \to (\pi+\tau,  -\phi)$ . 
\end{itemize}
Only the first of these isometries gives a priori a  `physically-allowed' 
orientifold of $AdS_3$ spacetime.
The last one is an involution of the hyperboloid,  but not of its
universal  covering space. The 
 second and third isometries 
change the orientation of time. Their fixed-point sets are  spacelike surfaces, 
whose interpretation is unclear. 

The `physical' orientifold is an $AdS_2$ slice dividing the 
$AdS_3$ cylinder in two equal parts. This meshes nicely with the fact that the only 
`physical' symmetric D-branes in  this model  have also $AdS_2$ geometry
\cite{BP}. An interesting question is whether one can  construct  
fully-consistent type I backgrounds as orientifolds of  type IIB theory on
$AdS_3$$\times $$S_3$. The possible orientifolds 
in orbifold spaces,  such as the BTZ black hole, may also be interesting to consider.

One of the motivation for studying the WZW models is 
that they arise in the near-horizon geometry  of  NS fivebranes.  
Recall that the throat region of $N=k+2$ parallel 
NS5-branes is described by an exact
conformal field theory \cite{CHS}
comprising: (i)  a level-k WZW 
model with group $SU(2)$  that corresponds to the 3-sphere surrounding
the branes, (ii) a Feigin-Fuchs or linear-dilaton field
corresponding to the radial direction, and (iii) six flat spacetime
coordinates along the fivebranes. Closed type II string theory in this
background is conjectured to be dual to the 5+1 dimensional 
worldvolume
theory of  the fivebranes. Since the coupling blows up at 
zero radius, classical string theory is, to be sure,  not reliable
in the  entire throat  region.

How are  our WZW orientifolds realized in this particular context?
Let us  take the  NS-fivebranes along the dimensions 056789, so that
the three-spheres  surrounding them  are given by
$X_1^2+X_2^2+X_3^2+X_4^2=L^2$, as  in section \ref{secgeom}.  
To preserve an $O(3)$ symmetry, the orientifold must intersect the branes
at right angles, and extend along either one or three transverse dimensions.
Assume for simplicity that the NS branes live  inside the orientifold,
i.e. that this latter extends also along 056789. There are then two 
distinct possibilities: (i) an ${\O}6$ plane cutting the three-sphere at the 
two poles, or (ii) an ${\O}8$ plane cutting $S^3$ at an equator two-sphere. 
These are precisely the orientifolds we  found in the WZW model.

Now it is a  well-known fact that when an ${\O}6$ crosses  an odd number of
NS-branes, it  changes its type from ${\O}6^-$ to $O6^+$ and vice versa 
\cite{e1,e2,orB2,orB3}. This agrees again with our observation that  when 
the Kac-Moody level  $k$
is odd, the north- and south-pole  orientifolds must be  of opposite type. 
The topological argument for this is, as a matter of fact, the same
in both contexts.

%%%%%%%%%%%%%%%%%%%%%%%%%%%%%%%%%%%%%%%%%%%%%%%%%%%%%

\acknowledgments 
We are grateful to C. Angelantonj,
E. Gimon, G. Pradisi, A. Sagnotti and A.
Schellekens for very useful conversations. C.B. thanks the Institute
for Theoretical Physics in Santa Barbara, and P.W. the CERN Theory
Division, for their kind hospitality at various stages of this work.
We also acknowledge the support of the European networks ``Superstring
Theory'' HPRN-CT-2000-00122 and ``The Quantum Structure of Spacetime''
HPRN-CT-2000-00131.

%%%%%%%%%%%%%%%%%%%%%%%%%%%%%%%%%%%%%%%%%%%%%%%%%%%%%

\appendix\setcounter{section}{1}\setcounter{equation}{0}
\section*{Appendix: Calculation of the $P$ matrix}

For the reader's convenience, we give here the computation of the $P$
matrix \cite{PSS}, equation \eqref{P}.\footnote{We thank Gianfranco
Pradisi for kindly communicating his notes.}

In order to simplify the formulae, the $SU(2)$ representations are
labelled by their dimension rather than by their spin. The matrices
$S$ and $T$ read:
\begin{equation}
\label{S}
S_{ab}=\sqrt{\frac{2}{k+2}}\; \sin \left(\frac{\pi ab}{k+2}\right)\ \ 
{\rm and }\ \  T_{ab}=  \displaystyle
             \exp\left(\frac{i\pi a^2}{2(k+2)}-\frac{i\pi}{4}\right)
             \delta_{ab}\ .
\end{equation}
It follows easily  that
\begin{equation}
(ST^2S)_{ab}=-\frac{i}{k+2}\;(L_{a-b}-L_{a+b})\ ,
\end{equation}
where
\begin{equation}
L_{x}=\sum_{c=0}^{k+1} \exp\left(\frac{i\pi  c^2}{k+2}\right)
                            \cos\left(\frac{\pi  xc }{k+2}\right).
\end{equation}
Now using  the following Gauss sum, valid for integer $x$:
\begin{equation}
\begin{array}{l} \displaystyle
\frac{1}{2}\sum_{c=0}^{k+1}
              \left[\exp\left(\frac{i\pi}{k+2}(c+{x\over 2})^2\right)
                   +\exp\left(\frac{i\pi}{k+2}(c-{x\over 2})^2\right)\right] \\
\displaystyle = \exp\left(\frac{i\pi}{4}\right)\sqrt{k+2}\;E_{k+x}
    + \exp\left(\frac{i\pi x^2}{4(k+2)}\right)O_{k+x}\ , 
\end{array}
\end{equation}
one finds after some algebra:
\begin{equation}
\begin{array}{ccl}
L_{x} = \displaystyle
        \exp\left(-\frac{i\pi x^2}{4(k+2)}+\frac{i\pi}{4}\right)
        \sqrt{k+2}\;E_{k+x} + O_{k+x}\ .
\end{array}
\end{equation}
Noting that   $E_{k+a+b}=E_{k+a-b}$ and $O_{k+a+b}=O_{k+a-b}$ leads to the equation 
\begin{equation}
\begin{array}{ccl}
L_{a-b}-\!L_{a+b} =  \displaystyle
      2\;\sqrt{k+2}\;\exp\left(\!-\frac{i\pi(a^2+b^2)}{4(k+2)}\right)
      \sin\left(\frac{\pi ab}{2(k+2)}\right)E_{k+a+b}\ .
\end{array}
\end{equation}
Multiplying \eqref{S} by $T^{1/2}$ on the left and right leads to 
the final  expression
for the  $P$ matrix, equation  \eqref{P}.

%%%%%%%%%%%%%%%%%%%%%%%%%%%%%%%%%%%%%%%%%%%%%%%%%%%%%


\begin{thebibliography}{99}


\bibitem{PSS}
G.~Pradisi, A.~Sagnotti and Y.~S.~Stanev,
``Planar duality in $SU(2)$ WZW models,''
Phys.\ Lett.\ B {\bf 354}, 279 (1995)
[hep-th/9503207].
%%CITATION = HEP-TH 9503207;%%

\bibitem{PSS2}
G.~Pradisi, A.~Sagnotti and Y.~S.~Stanev,
``The Open descendants of nondiagonal $SU(2)$ WZW models,''
Phys.\ Lett.\ B {\bf 356}, 230 (1995)
[hep-th/9506014]; 
``Completeness Conditions for
 Boundary Operators in 2D Conformal Field Theory,''
Phys.\ Lett.\ B {\bf 381}, 97 (1996)
[hep-th/9603097].
%%CITATION = HEP-TH 9603097;%%

\bibitem{SS}
A.~Sagnotti and Y.~S.~Stanev,
{`` Open Descendants in Conformal Field Theory},''
Fortsch.\ Phys.\ {\bf 44},  585 (1996) [hep-th/9605042].

%\cite{Fuchs:2000cm}
\bibitem{Sche}
J.~Fuchs, L.~R.~Huiszoon, A.~N.~Schellekens, C.~Schweigert and J.~Walcher,
``Boundaries, crosscaps and simple currents,''
Phys.\ Lett.\ B {\bf 495}, 427 (2000)
[hep-th/0007174].
%%CITATION = HEP-TH 0007174;%%

%\cite{BDS}
\bibitem{BDS}
C.~Bachas, M.~Douglas and C.~Schweigert,
``Flux stabilization of D-branes,''
JHEP {\bf 0005}, 048 (2000)
[hep-th/0003037].
%%CITATION = HEP-TH 0003037;%%


\bibitem{KS} C. Klimcik and P. Severa, {`` Open Strings and D-branes
in WZWN Model},''  Nucl. Phys. {\bf B383},  281 (1996)  [hep-th/9609112].

%\cite{AS}
\bibitem{AS}
A.~Y.~Alekseev and V.~Schomerus,
``D-branes in the WZW model,''
Phys.\ Rev.\  {\bf D60}, 061901 (1999)
[hep-th/9812193].
%%CITATION = HEP-TH 9812193;%%

\bibitem{S} S. Stanciu, {``D-branes in Group Manifolds},'' 
 { JHEP} {\bf 0001},  025 (2000) [hep-th/9909163].

\bibitem{ARS} 
A.Yu. Alekseev, A. Recknagel and  V. Schomerus, 
{``Non-commutative World-volume Geometries: Branes on $SU(2)$ and
Fuzzy Spheres},'' { JHEP} {\bf  9909},   023 (1999) [hep-th/9908040].

%\cite{Felder:2000ka}
\bibitem{FFFS}
G.~Felder, J.~Frohlich, J.~Fuchs and C.~Schweigert,
``The geometry of WZW branes,''
J.\ Geom.\ Phys.\  {\bf 34}, 162 (2000)
[hep-th/9909030].
%%CITATION = HEP-TH 9909030;%%

%\cite{Paw}
\bibitem{Paw}
J.~Pawelczyk,
``SU(2) WZW D-branes and their noncommutative geometry from DBI action,''
JHEP {\bf 0008}, 006 (2000)
[hep-th/0003057]. 

%\cite{Maldacena:2001ss}
\bibitem{MMS}
J.~Maldacena, G.~W.~Moore and N.~Seiberg,
``D-brane charges in five-brane backgrounds,''
JHEP {\bf 0110}, 005 (2001)
[hep-th/0108152].
%%CITATION = HEP-TH 0108152;%%


%\cite{Callan:1991at}
\bibitem{CHS}
C.~G.~Callan, J.~A.~Harvey and A.~Strominger,
``Supersymmetric string solitons,''
 hep-th/9112030.
%%CITATION = HEP-TH 9112030;%%

%\cite{Forste:1997ur}
\bibitem{Forste}
S.~Forste, D.~Ghoshal and S.~Panda,
``An orientifold of the solitonic fivebrane,''
Phys.\ Lett.\ B {\bf 411}, 46 (1997)
[hep-th/9706057].
%%CITATION = HEP-TH 9706057;%%


%\cite{Bianchi:1998gt}
\bibitem{BS}
M.~Bianchi and Y.~S.~Stanev,
``Open strings on the Neveu-Schwarz pentabrane,''
Nucl.\ Phys.\ B {\bf 523}, 193 (1998)
[hep-th/9711069].
%%CITATION = HEP-TH 9711069;%%

%\cite{Bachas:2001bt}
\bibitem{Bstr}
C.~Bachas,
``D-branes in some near-horizon geometries,''
arXiv:hep-th/0106234.
%%CITATION = HEP-TH 0106234;%%


\bibitem{GimP}
E.~G.~Gimon and J.~Polchinski,
``Consistency Conditions for Orientifolds and D-Manifolds,''
Phys.\ Rev.\ D {\bf 54}, 1667 (1996)
[hep-th/9601038].
%%CITATION = HEP-TH 9601038;%%


%\cite{Witten:1998xy}
\bibitem{orB1}
E.~Witten,
``Baryons and branes in anti de Sitter space,''
JHEP {\bf 9807}, 006 (1998)
[hep-th/9805112].
%%CITATION = HEP-TH 9805112;%%

%\cite{Evans:1997hk}
\bibitem{e1}
N.~Evans, C.~V.~Johnson and A.~D.~Shapere,
``Orientifolds, branes, and duality of 4D gauge theories,''
Nucl.\ Phys.\ B {\bf 505}, 251 (1997)
[hep-th/9703210].
%%CITATION = HEP-TH 9703210;%%

%\cite{Elitzur:1998ju}
\bibitem{e2}
S.~Elitzur, A.~Giveon, D.~Kutasov and D.~Tsabar,
``Branes, orientifolds and chiral gauge theories,''
Nucl.\ Phys.\ B {\bf 524}, 251 (1998)
[hep-th/9801020].
%%CITATION = HEP-TH 9801020;%%

%\cite{Hanany:2000fq}
\bibitem{orB2}
A.~Hanany and B.~Kol,
``On orientifolds, discrete torsion, branes and M theory,''
JHEP {\bf 0006}, 013 (2000)
[hep-th/0003025].
%%CITATION = HEP-TH 0003025;%%


%\cite{Bergman:2001rp}
\bibitem{orB3}
O.~Bergman, E.~Gimon and S.~Sugimoto,
``Orientifolds, RR torsion, and K-theory,''
hep-th/0103183.
%%CITATION = HEP-TH 0103183;%%

\bibitem{cou} N.~Couchoud, ``Branes and orientifolds of $SO(3)$'',
in preparation.

\bibitem{vilenkin}
N. Vilenkin, ``Special Functions and the Theory of Group
Representations'', American Mathematical Society (Providence 1968).

%\cite{Cardy:1989ir}
\bibitem{Ca}
J.~L.~Cardy,
``Boundary Conditions, Fusion Rules And The Verlinde Formula,''
Nucl.\ Phys.\  {\bf B324}, 581 (1989).
%%CITATION = NUPHA,B324,581;%%



%\cite{Hashimoto:2001xy}
\bibitem{DO4}
K.~Hashimoto and K.~Krasnov,
``D-brane solutions in non-commutative gauge theory on fuzzy sphere,''
Phys.\ Rev.\ D {\bf 64}, 046007 (2001)
[hep-th/0101145].
%%CITATION = HEP-TH 0101145;%%

%\cite{Hikida:2001py}
\bibitem{DO3}
Y.~Hikida, M.~Nozaki and Y.~Sugawara,
``Formation of spherical D2-brane from multiple D0-branes,''
arXiv:hep-th/0101211.
%%CITATION = HEP-TH 0101211;%%

%\cite{Pawelczyk:2001zi}
\bibitem{DO2}
J.~Pawelczyk and H.~Steinacker,
``Matrix description of D-branes on 3-spheres,''
arXiv:hep-th/0107265.
%%CITATION = HEP-TH 0107265;%%


%\cite{cc}
\bibitem{DO1}
D.~P.~Jatkar, G.~Mandal, S.~R.~Wadia and K.~P.~Yogendran,
``Matrix dynamics of fuzzy spheres,''
arXiv:hep-th/0110172.
%%CITATION = HEP-TH 0110172;%%

%\cite{Deser:1998rj}
\bibitem{DG}
S.~Deser and G.~W.~Gibbons,
%``Born-Infeld-Einstein actions?,''
Class.\ Quant.\ Grav.\  {\bf 15}, L35 (1998)
[hep-th/9803049].
%%CITATION = HEP-TH 9803049;%%


\bibitem{rev} J. Polchinski,
 ``{ TASI Lectures on D-branes},'' hep-th/9611050;
W. Taylor, ``{ Lectures on D-branes, Gauge Theory and M(atrices)},'' 
hep-th/9801182~;
C. Bachas, ``{ Lectures on D-branes},''  hep-th/9806199 . 

%\cite{Bianchi:1992eu}
\bibitem{BPS}
M.~Bianchi, G.~Pradisi and A.~Sagnotti,
``Toroidal compactification and symmetry breaking in open string theories,''
Nucl.\ Phys.\ B {\bf 376}, 365 (1992).
%%CITATION = NUPHA,B376,365;%%


%\cite{Bachas:2001fr}
\bibitem{BP}
C.~Bachas and M.~Petropoulos,
``Anti-de-Sitter D-branes,''
JHEP {\bf 0102}, 025 (2001)
[hep-th/0012234].
%%CITATION = HEP-TH 0012234;%%


%\cite{Brunner:2001fs}
\bibitem{Br}
I.~Brunner,
``On orientifolds of WZW models and their relation to geometry,''
arXiv:hep-th/0110219.
%%CITATION = HEP-TH 0110219;%%

\bibitem{HSS} L.R. Huiszoon, K. Schalm and  A.N. Schellekens,
``Geometry of WZW Orientifolds'', 
arXiv:hep-th/0110267. 

%%%%%%%%%%




\end{thebibliography}
\end{document}